# Unveiling the limits in the coherence of spin qubits against magnetic noise

*L. Escalera-Moreno,[a] A. Gaita-Ariño,[a,\*] and E. Coronado[a]*

[a]UIMM-ICMol, University of Valencia, c/ cat. José Beltrán 2, 46980 Paterna, Spain

Corresponding Author: * E-mail: alejandro.gaita@uv.es

Abstract: The realization of spin-based logical gates crucially depends on magnetically-coupled spin qubits. Thus, understanding decoherence when spin qubits are in close proximity will become a roadblock to overcome. Herein, we provide a general first-principles model that straightforwardly evaluates the spin bath effect on the qubit phase memory time $T_m$. The method is applied to a ground-spin $J=8$ magnetic molecule **1** displaying atomic clock transitions, which remarkably increase $T_m$ at unusually high spin concentrations. Besides reproducing experimental $T_m$ values calculated by recent models in simple spin-1/2 systems, our approach unveils the causes that limit the coherence reached at the clock transitions in more challenging systems such as **1**, where these previous models fail.

Spin qubits are promising candidates as building-blocks for storage and computation of quantum information.[1-4] These minimum-information units are encoded in energy levels of magnetic systems, and can be probed via different techniques such as paramagnetic resonance spectroscopy. As it happens in other physical qubits,[5,6] information stored in spin qubits is also affected by uncontrolled environmental interactions. The nature of this phenomenon –called quantum decoherence– depends on the qubit and on the experimental conditions. Two main mechanisms can collapse the quantum information saved in a spin qubit. These are vibrational decoherence,[7-10] and the magnetic noise caused by the interaction of the spin qubit with the nuclear and electron spin baths.[11]

The standard method employed to suppress this second mechanism consists in (i) placing qubit-carriers in nuclear-spin free environments and (ii) diluting these carriers in diamagnetic analogues.[12] Optimized combinations of these approximations remarkably increase the phase memory time $T_m$,[3] which is a figure of merit that characterizes how long a spin qubit can be kept in superposition of states. Nevertheless, isolation of qubits is impractical at the stage of device design, since the implementation of logical gates for quantum algorithms requires communication among close qubits.[13-15]

A strategy to overcome this drawback is operating at atomic clock transitions, also known as ZEFOZ (Zero First Order Zeeman) shifts, Fig.1.[16-19] These are avoided energy crossings in which Zeeman effect vanishes up to first order, making qubit coherence become remarkably insensitive to surrounding magnetic noise. Within molecule-based spin qubits, this method was recently demonstrated for the first time, and this allowed reaching long $T_m$ values at unusual high spin concentrations.[4] The need of working far from the high-dilution regime to allow inter-qubit operations demands to understand how qubit coherence is affected by a dense spin bath. The ultimate goal of building multipurpose architectures devoted to implement generic algorithms will require moving from ensemble experiments with randomly distributed spin qubits, to attaching magnetic entities on templated surfaces, forming spatially periodic arrays. There already exist some proposals, where spin qubits could be integrated in 2D arrays able to coherently control, read-out and, especially, mediate communication between a set of qubits.[20,21]

Herein, we aim to unveil the key factors limiting phase memory times in spin qubits displaying ZEFOZ shifts. The target system is a single-crystal of a Ho$^{III}$-based spin qubit **1**, Fig.2.[4]. Four peculiar narrow regions appear in the $T_m$ magnetic field dependence of **1**, where the phase memory time sharply increases up to a maximum value, Fig.3. To understand the origin of this limiting value, we firstly applied a recent model that satisfactorily explained the $T_m$ evolution at increasing concentration of a spin-1/2 system **2**, Fig.2.[22] This model generalizes the van Vleck second moment expression to an anisotropic Landé factor. However, it overestimates the $T_m$ top values reached at the ZEFOZ shifts in **1** as we will see below. This motivates us to develop a new *ab initio* model to quantify both the nuclear and electron spin bath effects on $T_m$. Our method is general and successfully reproduces the right height of $T_m$ in **1**. Besides, it also recovers the experimental $T_m$ trend in **2** as a function of spin concentration.

The starting point of our model is a decoherence rate $\gamma$ defined as $\gamma = 2\hbar / T_m \Delta$,[11,23-25] being $\Delta$ the gap energy between the two spin states of the qubit and $T_m$ the phase memory time. We incorporate the contribution to $T_m$ from the nuclear and electron spin baths. Thus, we need two rates: $\gamma_n$ for the nuclear bath and $\gamma_e$ for the electron bath. Assuming additive rates,[11,25] the collective phase memory time is $T_m^{n+e} = 2\hbar / \left[ \left( \gamma_n + \gamma_e \right) \Delta \right]$. The next step is to relate $\gamma_n$ and $\gamma_e$ with the nuclear $E_n$ and electron $E_e$ contributions to the echo line half-width. Under the so-called high-field regime, which means $E_n \ll \Delta$ and $E_e \ll \Delta$, these rates can be calculated perturbatively,[11,24,25] obtaining $\gamma_n = 2 \left( E_n / \Delta \right)^2$ and $\gamma_e = 2 \left( E_e / \Delta \right)^2$. The range of nuclear and electron spin concentrations often lies inside this regime,[11,22,25], leading to $T_m^{n+e} = \hbar \Delta / \left( E_n^2 + E_e^2 \right)$. The main goal is now to calculate $E_n$ and $E_e$ in terms of the nuclear and electron spin positions in each bath.

The $E_n$ expression, accounting for magnetic nuclei precessional motions that dephase qubit dynamics, was derived elsewhere.[25] Whenever spin diffusion is the dominant nuclear decoherence source, one can switch into the model that resulted in quantitative agreements with experimental coherence decays when applied to semiconductor and molecular quantum architectures.[26,27] Our expression for $E_e$ is inspired on the abovementioned precessionally-driven nuclear decoherence model,[25] and reads as follows:

$$E_e^2 = \frac{1}{N} \sum_{j=1}^{N-1} \sum_{k>j}^{N} \left\| \vec{E}_{jk} \right\|^2 \quad (1)$$

Eq. (1) determines the total square contribution to the echo line half-width from the electron spin bath. $N$ is the number of electron spins in the bath, and $\left\| \vec{E}_{jk} \right\|^2$ accounts for the pair-wise square contribution of the $j-k$ electron spin interaction:

$$\left\| \vec{E}_{jk} \right\|^2 = \left( \frac{\mu_0 \mu_B^2}{8\pi r_{jk}^3} \right)^2 \sum_{\alpha=x,y,z} \left[ g_\alpha \langle \hat{J}_\alpha \rangle \left( \left( 1 - 3 \left( \frac{(\vec{r}_{jk})_\alpha}{r_{jk}} \right)^2 \right) g_\alpha \langle \hat{J}_\alpha \rangle - 3 \frac{(\vec{r}_{jk})_\alpha}{r_{jk}^2} \sum_{\substack{\beta=x,y,z \\ \beta \neq \alpha}} (\vec{r}_{jk})_\beta g_\beta \langle \hat{J}_\beta \rangle \right) \right]^2$$

(2)

The Boltzmann-averaged expectation values $\{\langle \hat{J}_\alpha \rangle\}_{\alpha=x,y,z}$ of the total angular momentum operator arise from considering an electron spin bath in a thermalized state at a temperature $T$:

$$\left\{ \langle \hat{J}_\alpha \rangle = \left( \sum_{k=1}^{(2J+1)(2I+1)} e^{-E_k/k_B T} \langle \hat{J}_\alpha^2 \rangle_k \Big/ \sum_{k=1}^{(2J+1)(2I+1)} e^{-E_k/k_B T} \right)^{1/2} \right\}_{\alpha=x,y,z}$$

(3)

We consider that each magnetic center is described by the same spin Hamiltonian. Thus, Landé factors $\{g_\alpha\}_{\alpha=x,y,z}$, expectation values $\{\langle \hat{J}_\alpha \rangle\}_{\alpha=x,y,z}$, and energies $E_k$ are conserved from each center to any other one. This approximation is often encountered in magnetically-doped crystals, and in 2D systems such as **2**.[22] Nevertheless, the model can be readily expanded to incorporate distributions in Hamiltonian parameters. In our analysis, we include in the spin Hamiltonian the hyperfine coupling between the electron and the nuclear spins of the magnetic center. Thus, we obtain $(2J+1)(2I+1)$ energies and wave-functions, being $J$ and $I$ the electron and nuclear total angular momentum quantum numbers. Details on obtaining Eq. (2) are found in SI.

Eventually, the hybridization of the magnetic ion orbitals with the ligands may lead to a decrease in the electron spin magnitude.[22] This effect is described by a covalence parameter $0 < \varepsilon \leq 1$ .[28] The maximum value $\varepsilon = 1$ corresponds to ionic bonds. We incorporate $\varepsilon$ as an effective correction to $E_n$ and $E_e$. Thus, our phase memory time reads as $T_m^{n+e} = \hbar \Delta / \left[ \varepsilon^2 \left( E_n^2 + E_e^2 \right) \right]$.

The system **1**, HoW$_{10}$, is described by $\hat{H} = \sum_{k=2,4,6} B_k^0 \hat{O}_k^0 + B_4^4 \hat{O}_4^4 + \vec{I} \cdot \mathbf{A} \cdot \vec{J} + \mu_B \vec{B} \cdot \mathbf{g} \cdot \vec{J}$, with ground electron and nuclear spins $J = 8$ and $I = 7/2$.[4] The extended Stevens operators $\hat{O}_2^0$, $\hat{O}_4^0$, $\hat{O}_6^0$, $\hat{O}_4^4$,[29] account for the crystal field, with $B_2^0 = 0.601$ cm$^{-1}$, $B_4^0 = 6.96 \cdot 10^{-3}$ cm$^{-1}$, $B_6^0 = -5.10 \cdot 10^{-5}$ cm$^{-1}$, $B_4^4 = 3.14 \cdot 10^{-3}$ cm$^{-1}$.[4] The hyperfine coupling and Zeeman parameters $A_z = 830$ MHz, $g_z = 1.25$ characterize the molecular easy axis and the magnetic field direction, $\vec{B}$. When combined with the nuclear spin projections, the ground doublet $m_J = \pm 4$ gives rise to a low-lying manifold of 16 states, Fig.3.[30] The sizeable interaction $B_4^4$ generates an energy gap $\Delta \sim 9.18$ GHz at the ZEFOZ shifts, which defines the long-lived spin qubit.[4] Lanthanide f-orbitals are very internal and do not deploy a significant spin density towards the ligands, thus $\varepsilon = 1$.

In Fig.3, the qubit Zeeman curves have a slope ca. zero at the ZEFOZ fields, making $\Delta$ become insensitive to magnetic noise up to first order.[4] This remarkably increases $T_m$, but only up to a maximum value. Nuclear decoherence is negligible, as the calculation $T_m^n \sim 300$ μs is much above the experimental values. Instead, our $T_m^e$ calculation, free of fitting parameters, excellently agrees with the maximum $T_m$ values at the relevant concentrations, which span over one order of magnitude, Fig.3. The high-field regime holds as the highest $E_e \sim 0.05$ GHz $\ll \Delta \sim 9.18$ GHz (10% conc.). Since the calculated expectation values $(3)$, $\langle \hat{J}_x \rangle = \langle \hat{J}_y \rangle = 5.3$, $\langle \hat{J}_z \rangle = 4.0$, remain constant with $|\vec{B}|$ and are non-zero, $E_e^2$ takes a strictly positive value. Hence, $T_m^e \propto 1/E_e^2$ cannot diverge and reach an arbitrarily high value.

Out of the ZEFOZ fields one can note the appearance of three small peaks in the calculated $T_m^e$ evolution of Fig.3. As $T_m^e \propto \Delta / E_e^2$ and $E_e^2$ does not change with $|\vec{B}|$, these peaks arise because of the small increase in $\Delta$. Our model incorporates in $T_m^e$ the effect of a thermal electron spin bath, which might be the cause limiting qubit coherence at the ZEFOZ fields in the spirit of the match between experimental and calculated $T_m$ values. Out of these fields other decoherence sources, only suppressed at the clock transitions, might be the limiting mechanism and cause the fast decay in $T_m$. Our approach does not account for these extra mechanisms, which will be determined in a future work, and cannot recover the $T_m$ decay.

Whenever electron decoherence dominates, tuning $T_m^e$ needs to properly engineer $\Delta$ and $E_e^2$. The former depends on the electronic structure, and the latter depends additionally on magnetic dilution. The gap $\Delta$ in **1** at the ZEFOZ fields is set by the $B_4^4$ parameter, which is activated because of the deviation from the $D_{4d}$ symmetry of the Ho-coordinating oxygen atoms set.[4] Simple calculations reveal that $\Delta$ scales with $B_4^4$ as $\Delta$ (GHz) = $2.0 B_4^4$ (cm$^{-1}$) + $926900 \left( B_4^4 \right)^2$ (cm$^{-2}$). Besides, the expectation values $(3)$ are unaffected by changing $B_4^4$ in a wide range around $3.14 \cdot 10^{-3}$ cm$^{-1}$. Thus, given a spin concentration, we expect that a rise in $B_4^4$ will increase $\Delta$ while keeping $E_e^2$ unaltered. Since $T_m^e \propto \Delta / E_e^2$, the phase memory should be consequently increased. To understand how $T_m^e$ scales with magnetic dilution, let us fix the electronic structure with given values of $\Delta$, $g_\alpha$ and Eq. $(3)$. If we replace $r_{jk}$ in Eq. $(2)$ by an average effective distance, $E_e^2$ becomes proportional to $(1/N) N(N-1)/2$. Since $N$ is large, $T_m^e \propto 1/N$. To assess the validity of this expression, let us recall that $T_m$ in **1** is ten-fold larger as electron spin concentration decreases by one order of magnitude. Thus, we expect $N_{10\%} / N_{1\%} = 10$. Indeed, the $E_e^2$ calculation is converged with $N_{10\%} = 88259$, $N_{1\%} = 8832$, see SI, and $N_{10\%} / N_{1\%} = 9.993$. In summary, our model successfully reproduces the $T_m$ top values at the ZEFOZ fields in **1**, and provides insight on the factors limiting qubit coherence (see also below).

The method herein developed also matches the $T_m$ experimental values recovered by previous models in simple spin-1/2 systems. Indeed, as a further check we also applied our model to **2**, CuPc.[22] This is a $J = S = 1/2$ magnetic system complemented with $I = 3/2$. The energy level scheme is composed of eight spin states, which arise from the spin Hamiltonian $\hat{H} = \vec{\hat{I}} \cdot A \cdot \vec{\hat{J}} + \mu_B \vec{B} \cdot g \cdot \vec{\hat{J}}$. The copper electron-nuclear hyperfine coupling is set by $A_z = -648$ MHz, $A_x = A_y = -83$ MHz, and the Zeeman effect is due to an external field $|\vec{B}| = 311.5$ mT, with $g_z = 2.1577$, $g_x = g_y = 2.0390$. The transition $|m_J = -1/2, m_I = -1/2\rangle \rightarrow |m_J = +1/2, m_I = -1/2\rangle$ defines the qubit, with a gap $\Delta \sim 9.73$ GHz.

The main contribution from the nuclear spin bath in **2** is due to the four Cu-coordinating nitrogen nuclei in Fig.S1.[22] This interaction corresponds to a contact hyperfine coupling, see SI, much stronger than a magnetic dipolar interaction, with coupling constants $A_{xx}^N = 57$ MHz, $A_{yy}^N = A_{zz}^N = 45$ MHz.[22] By setting $\varepsilon = 0.74$,[22] we get $T_m^n = 2.1$ μs, similar to 2.2 μs as calculated elsewhere.[22] We consider that nitrogen nuclei have spins $I_N = 1$, since this is the most occurring isotope. The high-field approximation holds as $E_n \sim 0.03$ GHz $\ll \Delta \sim 9.73$ GHz. Calculating $E_e$ requires the positions of the electron spins in the bath, limited to a spherical granule of 50 nm diameter,[22] see Fig.S1 and SI. The high-field approximation holds again as the highest $E_e \sim 0.06$ GHz $\ll \Delta \sim 9.73$ GHz (10% conc.), and the calculated nuclear-electron $T_m$ values with $\varepsilon = 0.74$ satisfactorily agree with the experiment, Fig.4.

To end up, it is interesting to unveil additional factors that can make $T_m$ values be different between protected systems via ZEFOZ shifts (**1**) and simple $S = 1/2$ systems (**2**). Indeed, at the highest electron spin concentration, the $T_m^e$ values in **1** at the ZEFOZ fields are appreciably higher than that of **2**, Figs.3,4. To figure this difference out, let us fix an 8.2% conc. in **1**, equivalent to 10% in **2**, see SI. The Eq.(3) values in **2** are $\langle \hat{J}_x \rangle = \langle \hat{J}_y \rangle = \langle \hat{J}_z \rangle = 0.5$, much smaller than those of **1**. As $E_e^2$ is initially proportional to $\{\langle \hat{J}_\alpha \rangle\}_{\alpha=x,y,z}$, we would expect a greater $T_m^e$ value in **2**. To explain the rather opposite behavior, we need to focus on the Landé tensor of **1** and **2**. While all the Landé factors $\{g_\alpha\}_{\alpha=x,y,z}$ are non-zero in **2**, only $g_z$ is different from zero in **1**. Since Eq.(2) is a three-term sum proportional to $g_\alpha$, $E_e^2$ takes a smaller value for **1** which results in a larger $T_m^e$ value. Indeed, simple calculations show that increasing $g_x$, $g_y$ values in **1** decreases $T_m^e$, which reveals the crucial role of having a highly axial Landé tensor. In the case of **2**, lowering $g_x = g_y = 2.039$ increases $T_m^e = 0.44$ μs up to a maximum value of 0.96 μs at $g_x = g_y = 0.750$. Thus, depending on the system, fine tuning of $g_\alpha$ may be required to maximize $T_m$.

Herein, we have developed a straightforward *ab initio* model that satisfactorily simulates the influence of magnetic noise on the phase memory time $T_m$ of a given spin qubit. The model works in wide spin concentration ranges, and incorporates contributions from nuclear and electron spin baths. It is also valid in isotropic and anisotropic magnetic systems, and can deal with distributions in electron structure and Zeeman parameters. Besides being successful in simple spin-1/2 systems, it accounts for qubit coherence in challenging systems displaying ZEFOZ shifts, where previous models fail. Indeed, our results state that a properly engineered electronic structure can result in enhanced ZEFOZ coherences. Namely, special focus should be put on highly-axial Landé tensors and key parameters that control the qubit energy gap. Thus, this general method constitutes a widely applicable tool able to offer insight on understanding decoherence towards integrating spin qubits in quantum devices. Because of the potential applicability of ZEFOZ-based approaches in providing close-proximity and coherent spin qubits, in a future work we will address the issue of the fast $T_m$ decay out of the ZEFOZ fields.

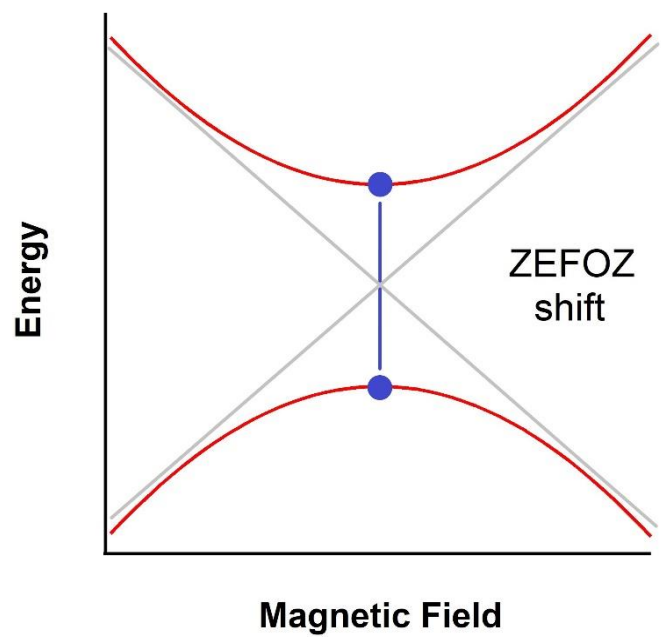

**Figure 1**. Schematic representation of a ZEFOZ (Zero First Order Zeeman) shift in blue at an avoided crossing between two energy levels in red. Note the vanishing slopes at the ZEFOF field.

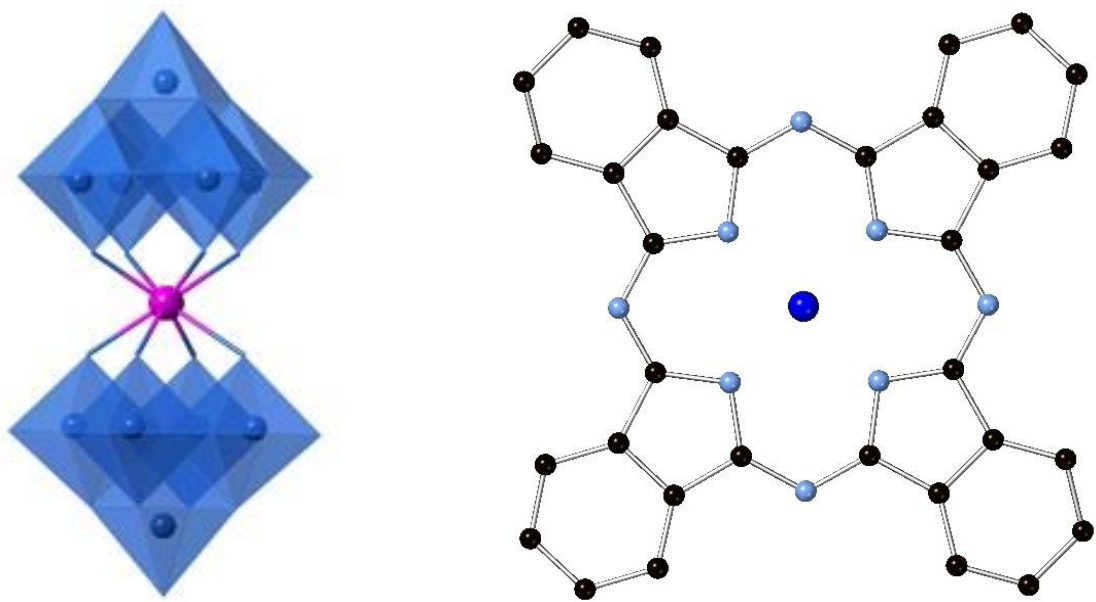

**Figure 2**. Left: **1**; blue spheres: tungsten, magenta sphere: holmium, polyhedron vertexes: oxygen. Right: **2**; black: carbon, light blue: nitrogen, dark blue: copper. Hydrogen atoms in **2** are omitted for clarity.

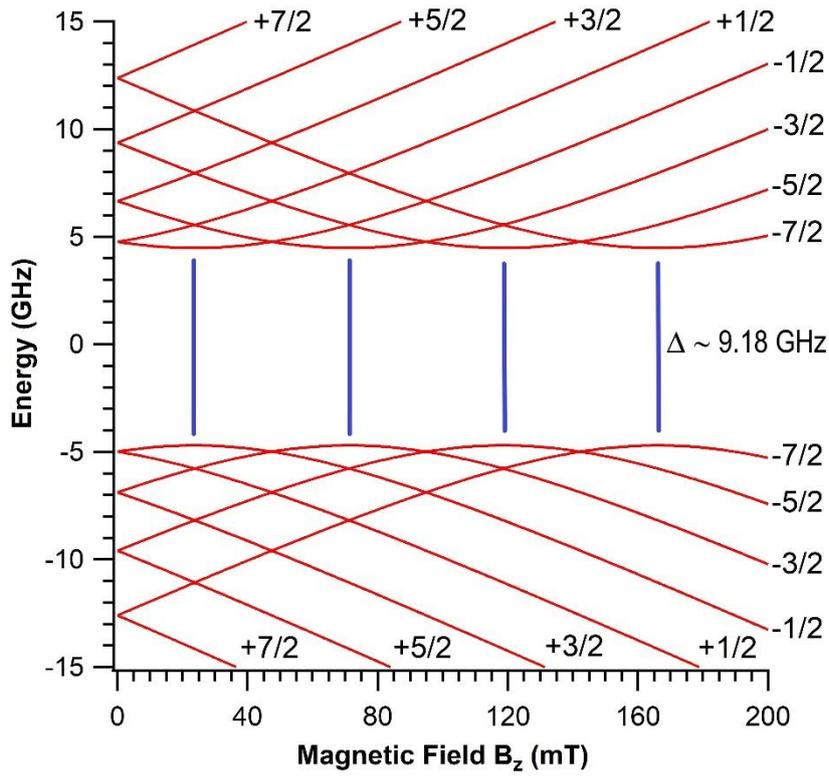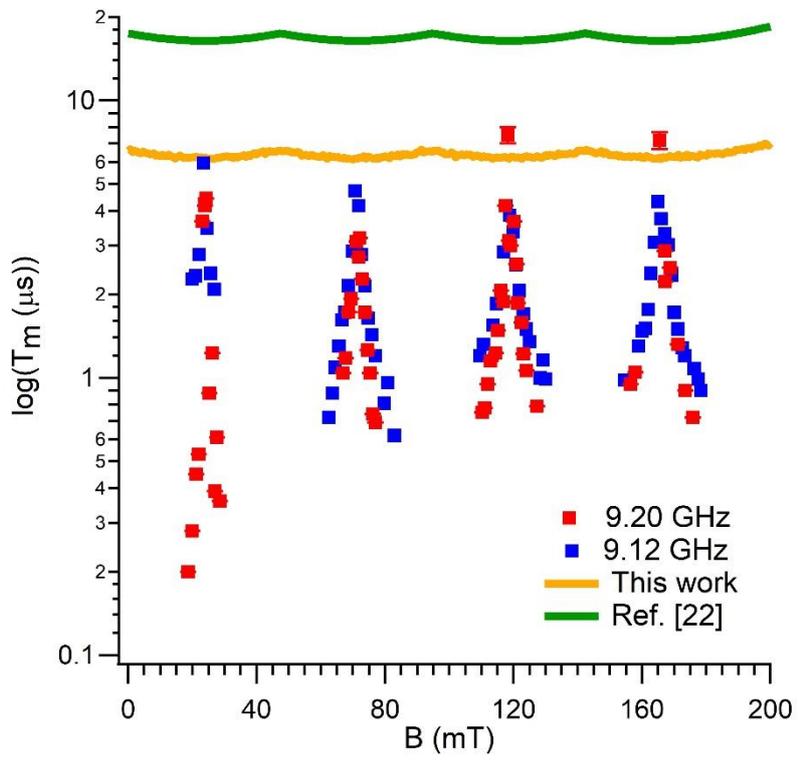

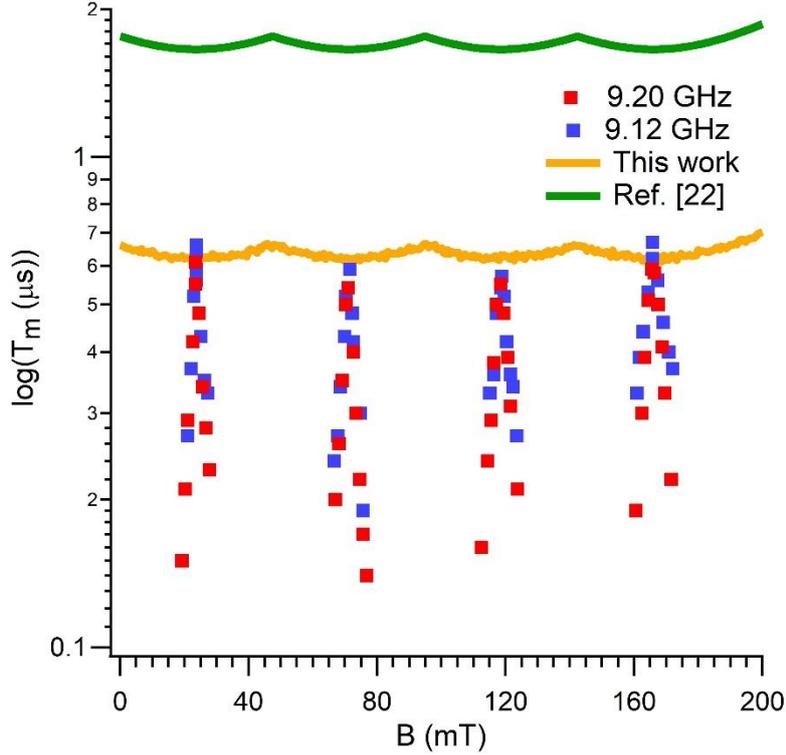

**Figure 3**. Top: Zeeman energy evolutions of **1** labeled with the $m_I$ projections. Blue lines are ZEFOZ shifts. Middle and Bottom: magnetic field dependence of $T_m$ in **1** at $T = 5K$ for two electron spin concentrations. Middle: 1%. Bottom: 10%. Blue and red points are experimental $T_m$ values at different microwave frequencies.[4] Green and orange curves are theoretical calculations using the model in ref.[22] and that of the present work, resp.

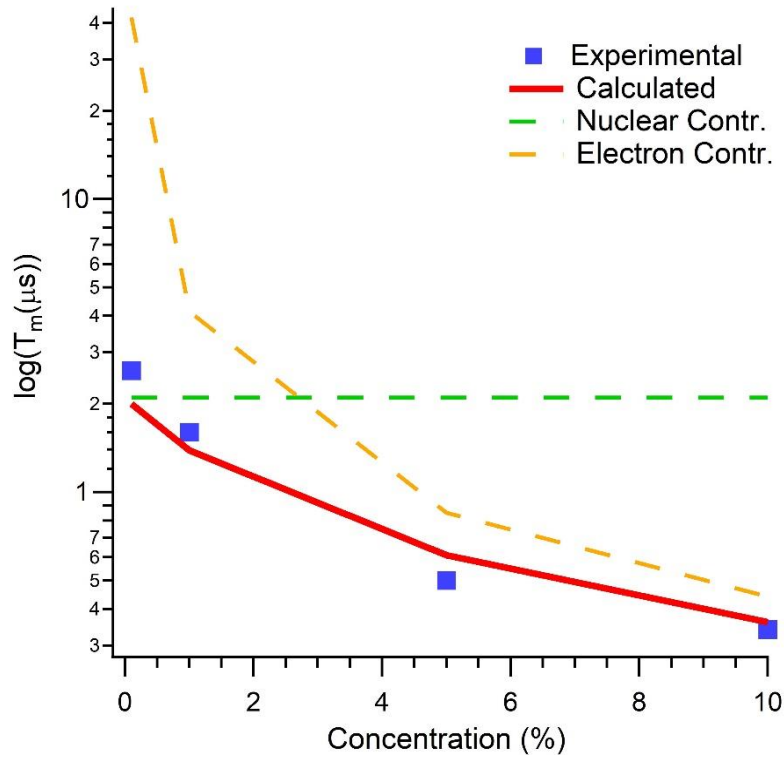

**Figure 4**. Phase memory time evolution of **2** at $T = 5K$ with the electron spin concentration. The experimental blue points are from ref.[22]. Green and orange dashed lines are our calculated nuclear and electron spin bath contributions to $T_m$ (red curve).

Supporting Information

-Supplementary Information

-CrystalMaker file ready to generate the CuPc electron spin bath: CuPc-CM

-CrystalMaker file ready to generate the HoW$_{10}$ electron spin bath: HoW10-CM

Acknowledgements

The present work has been funded by the EU (COST Action CA15128 MOLSPIN, ERC-2014-CoG-647301 DECRESIM, ERC-2018-AdG-788222 MOL-2D), the Spanish MINECO (Unit of excellence "María de Maeztu" MDM-2015-0538 and grants MAT2017-89993-R, CTQ2017-89528-P) and the Generalitat Valenciana (Prometeo Program of excellence). L.E.-M. thanks also the Generalitat Valenciana for a VALi+D predoctoral contract and A.G.-A. thanks funding by the MINECO (Ramón y Cajal contract). We also thank Dr. I. S. Tupitsyn for fruitful discussions.